\documentclass{aa}  

\usepackage{natbib,twoopt}
\usepackage[breaklinks=true]{hyperref} 
\bibpunct{(}{)}{;}{a}{}{,}             
\usepackage{graphicx}
\usepackage{txfonts}
\usepackage{hyperref}
\hypersetup{
    colorlinks=true,
    linkcolor=blue,
    citecolor = blue,
    filecolor=magenta,      
    urlcolor=blue}
\usepackage{environ}
\NewEnviron{myequation}{%
\begin{equation}
\scalebox{0.8}{$\BODY$}
\end{equation}}

\graphicspath{{./}{Figures/}}

\defcitealias{alvarez_22}{AG22}

\authorrunning{Alvarez Garay et al.}
\titlerunning{MgAl burning chain in $\omega$ Centauri}

\begin{document} 

\title{MgAl burning chain in $\omega$ Centauri}

\author{Deimer Antonio Alvarez Garay \inst{1}\fnmsep\inst{2},
    Alessio Mucciarelli \inst{1}\fnmsep\inst{2}, 
    Michele Bellazzini \inst{2},
    Carmela Lardo \inst{1},
    \and Paolo Ventura \inst{3}
          }

\institute{Dipartimento di Fisica e Astronomia, Università degli Studi di Bologna, Via Gobetti 93/2, I-40129 Bologna, Italy\\
\email{deimer.alvarezgaray2@unibo.it}
    \and
    INAF, Osservatorio di Astrofisica e Scienza dello Spazio di Bologna, Via Gobetti 93/3, I-40129 Bologna, Italy
    \and
    INAF, Osservatorio Astronomico di Roma, Via Frascati 33, I-00040 Monte Porzio Catone, Roma, Italy}



\abstract
{In this study, we report the results of Fe, Mg, Al, and Si abundances analysis for a sample of 439 stars in $\omega$ Centauri, using high-resolution spectra obtained with the VLT/FLAMES multi-object spectrograph. Our analysis reveals the presence of four distinct Fe populations, with the main peak occurring at low metallicity, consistent with previous literature findings. We observe a discrete and pronounced Mg-Al anti-correlation, which exhibits variations in shape and extension as a function of metallicity. Specifically, this anti-correlation is present in stars with metallicities lower than approximately $-1.3$ dex, while it becomes less evident or absent for higher [Fe/H] values. Additionally, we detect (anti-) correlations between Mg and Si, and between Al and Si, whose extensions also vary with metallicity, similar to the Mg-Al anti-correlation. These results suggest that the MgAl cycle plays a crucial role in the formation of multiple populations in $\omega$ Centauri, with the presence of all (anti-) correlations at metallicities lower than -1.3 dex providing evidence for the burning of Mg at very high temperatures ($> 10^8$ K), at least in the metal-poor regime. Furthermore, we observe a clear trend of stars with [Al/Fe] $> +0.5$ dex as a function of metallicity, confirming for the first time the existence of the two channels of Al production and destruction. This evidence can help to provide further constraints on the potential nature of the polluters responsible for the observed chemical anomalies in this stellar system. Finally, we find that the two most metal-poor populations identified in our sample are compatible with null or very small metallicity dispersion and we discuss how this result fit into a scenario where $\omega$ Centauri is the remnant of a disrupted nucleated dwarf galaxy.
}


\maketitle
%
\section{Introduction}
The vast spectroscopic and photometric evidence obtained over the past three decades has conclusively shown that globular clusters (GCs) host multiple populations of stars with prominent variations in the abundance of light elements (C, N, O, Na, Mg, Al; \citealt{carretta09}; \citealt{meszaros15}; \citealt{pancino17}; \citealt{masseron_19}). Previously, the observed large intrinsic spreads in light elements were mainly thought to be associated with significant variations in iron abundance only in a small subset of the entire globular cluster population (i.e. $\omega$ Centauri, M54, Terzan5, and Liller1; \citealt{NDC95}; \citealt{sarajedini_95}; \citealt{ferraro_09}; \citealt{crociati_23}). However, recent photometric and spectroscopic studies have found evidence suggesting that small-to-moderate iron spreads (from less than 0.05 up to 0.3 dex) may be relatively common in massive clusters (\citealt{legnardi_22}; \citealt{lardo_22}; \citealt{lardo_23}; \citealt{lee_22}; \citealt{lee_23}; \citealt{monty_23}).\\
All the observed chemical differences are structured in well-defined patterns, such as the C-N, Na-O, and Mg-Al anti-correlations (see, e.g., \citealt{gratton_04}; \citealt{gratton_12}; \citealt{gratton_19};  \citealt{carretta09}; \citealt{pancino17}; \citealt{meszaros_20}). The overall observational evidence is interpreted as the characteristic signature of self-enrichment within clusters, where low-velocity material processed through the hot CNO cycle and its secondary NeNa and MgAl chains (e.g., \citealt{langer_93}; \citealt{prantzos_07}) is incorporated in a subsequent generation of stars. Indeed, the majority of theoretical models for the formation of multiple populations involve the occurrence of two or more episodes (in some clusters only two main populations are detected) of star formation where CNO-enriched stars (second population, 2P) were formed out of matter polluted by massive stars with field-like composition (first population, 1P) within the first 100-200 Myr of the cluster life. \\ 
A number of polluters were proposed in the literature, including intermediate-mass stars in their asymptotic giant branch (AGB) phase (\citealt{dercole_10}), fast rotating massive stars (FRMS; \citealt{krause_13}), novae (\citealt{maccarone_12}; \citealt{denissenkov_14}), interacting binary stars (\citealt{mink_09}) and supermassive stars (\citealt{denissenkov_hart_14}). Nonetheless, all self-enrichment models put forward so far fail to reproduce the observed chemical anti-correlations and the number ratio between 1P and 2P stars (see e.g . \citealt{bastian_18} and references therein for a discussion).  \\
The Mg-Al anti-correlation is of special relevance in this context because, in contrast to other anti-correlations (such the C-N and Na-O ones), its extension differs significantly from one cluster to another and is absent in some GCs (\citealt{meszaros15}). These two elements are involved in the hot MgAl cycle, which works at temperatures higher than those of the CNO and NeNa cycles ($\gtrsim 10^8$ K; \citealt{ventura_16}). As a result, the analysis of these two elements can place substantial constraints on the nature of polluters responsible for the chemistry seen in GCs. Additionally, compared to the other elements involved in chemical anomalies (C, N, O, and Na; \citealt{denisenkov_90}), Mg and Al are not affected by deep mixing processes occurring during the red giant branch phase; therefore, their chemical abundances reflect the initial chemical composition of the gas from which stars formed. \\\\
$\omega$ Centauri (NGC 5139), is a highly complex stellar cluster, usually classified as a GC according to its morphology and mass. It is the most massive among GCs with a mass of $(3.94\pm0.02)\cdot 10^6\;\mathrm{M_{\odot}}$ (\citealt{baumgardt_18}); it spans a wide metallicity range ($-2.2\lesssim$ [Fe/H] $\lesssim -0.5$ dex)  with at least four main populations with different iron content (\citealt{NDC95}; \citealt{pancino_02}; \citealt{johnson_10}; \citealt{marino_11}). The multi-modal iron distribution seen in $\omega$ Centauri suggests that this system underwent multiple star formation events lasting a few Gyr (\citealt{smith_00}; \citealt{sollima_05_age}; \citealt{romano_10}; \citealt{villanova_14}), at variance with the genuine GCs. The prevalent interpretation is that  $\omega$ Centauri is the remnant of an old nucleated dwarf galaxy that the Milky Way accreted in the past (\citealt{bekki_03}). $\omega$ Centauri also exhibits large star-to-star variations in light elements, that manifest as correlations/anti-correlations (\citealt{NDC95}; \citealt{smith_00}; \citealt{johnson_10}; \citealt{marino_11}; \citealt{meszaros_21}).\\
$\omega$ Centauri is the only GC-like system showing all the anti-correlations usually observed in (some or all) genuine GCs, i.e. Na-O (\citealt{johnson_10}; \citealt{marino_11}), Na-Li (\citealt{mucciarelli_18_Li}), Mg-Al (\citealt{NDC95}; \citealt{smith_00}; \citealt{meszaros_21}), Mg-Si and Mg-K (\citealt{meszaros_20}; \citealt{alvarez_22}). All of these chemical anomalies point to the extreme NeNa and MgAl chains playing a crucial role in the formation of multiple populations in  $\omega$ Centauri. In particular, proton capture reactions operating at temperatures higher than $10^8$ K could explain the anti-correlations between Mg and Si and Mg and K. In the case of Mg-Si anti-correlation part of the previous synthesized Al is used to produce Si, whereas proton capture on Ar nuclei leads to the synthesis of K in the Mg-K anti-correlation. \\
Using the same sample analyzed in \citet{alvarez_22} (hereafter \citetalias{alvarez_22}), in this study we provide the chemical abundances of those elements (Mg, Al, and Si) participating in the MgAl cycle in $\omega$ Centauri for a total of 439 member stars along the Red Giant Branch (RGB). The structure of this paper is as follows: the data are presented in Section $\ref{sec:observations}$, the chemical analysis is detailed in Section $\ref{sec:analysis}$, the metallicity distribution and Mg-Al-Si abundance variations are illustrated and discussed in Section $\ref{results}$, in Section \ref{comparison_meszaros} is performed a comparison with the analysis done by \citet{meszaros_21}, and findings and conclusions are summarized in Section $\ref{discussion}$.

\section{Observations and atmospheric parameters} \label{sec:observations}
\begin{figure}
   \centering
   \includegraphics[width=9 cm]{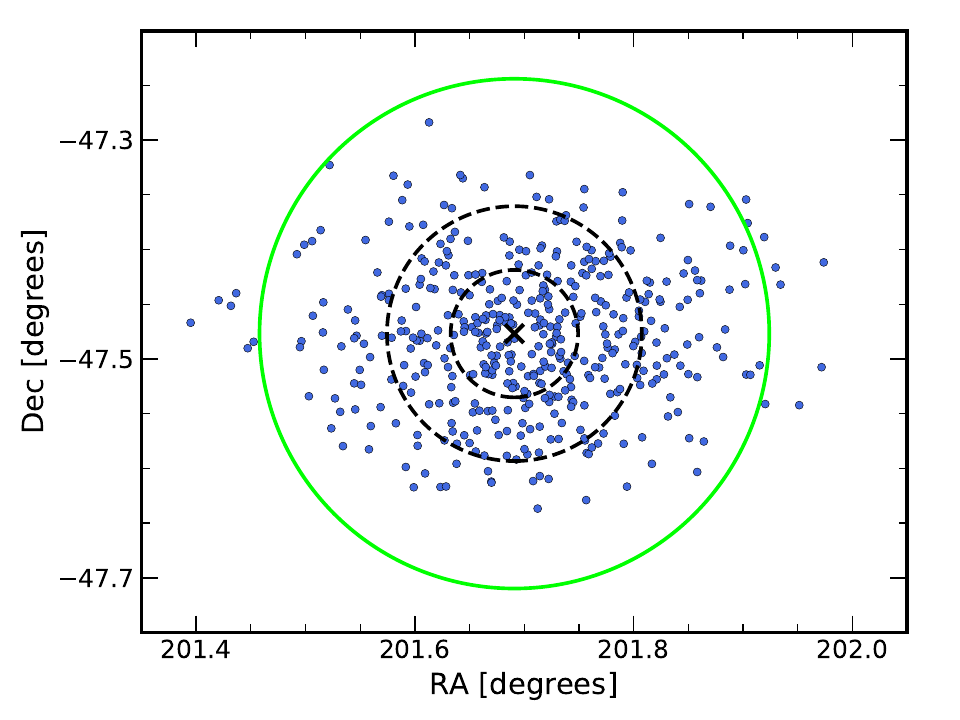}
      \caption{The coordinate positions of FLAMES targets are displayed by the red circles. The black cross denotes the cluster center ($-201\rlap{.}^\circ6910, -47\rlap{.}^\circ4769$) according to \citet{vanleeuwen_00}. The dashed black circles show 2.5 and 5 times the core radius ($r_c = 1\rlap{.}^{'}40$; \citealt{harris_96}). 424 out of the 439 stars that were analyzed in this study are contained within the green circle, which is 10 times the core radius.}
         \label{fig:distrib_rad}
\end{figure}

This study is a continuation of our recent work (\citetalias{alvarez_22}) in which we analyzed the extension of the Mg-K anti-correlation among the RGB stars of $\omega$ Centauri (Figure 1 in \citetalias{alvarez_22} shows the position in the color-magnitude diagram of the considered stars). The data-set is the same used in \citetalias{alvarez_22}, and consists of high-resolution spectra collected with the multiobject spectrograph FLAMES (\citealt{pasq_02}), mounted on UT2 (Kueyen) at the ESO-VLT Observatory in Cerro Paranal, within the ESO program 095.D-0539 (P.I. Mucciarelli). The observations were performed in the GIRAFFE mode that allows to allocate simultaneously up to 132 fibers. The adopted setups are HR11 (with a wavelength range from 5597 to 5840 $\AA$ and a spectral resolution of 29500) and HR18 (with a wavelength range from 7648 to 7889 $\AA$ and a spectral resolution of 20150). The first setup allows us to measure up to five lines of Si, while the second Al doublet at 7835 and 7836 $\AA$. \\
All the observed targets are on the RGB of $\omega$ Centauri, and their membership was confirmed by multiple literature sources (\citealt{NDC95}; \citealt{johnson_10}; \citealt{marino_11}). A total of 439 stars were analyzed: 345 of them are in common with \citet{johnson_10}, 82 with \citet{marino_11}, and 12 with \citet{NDC95}. Further, we considered only stars that are not contaminated by neighbor stars within the GIRAFFE fibers size.\\
Four configurations of targets were defined and each of them was observed with both HR11 and HR18 setups. Due to the brightness of the targets ($10.7<G<14.5$), for each configuration two exposures of 1300 s and two of 300 s each were sufficient to reach a signal-to-noise ratio (S/N) $\sim 70$ and S/N $\sim 100$ for HR11 and HR18, respectively. The observation of each configuration was split in two exposures in order to get rid of the effects of cosmic rays and other transient effects. Finally, during each exposure about 15 fibers were dedicated to observe empty sky regions in order to sample the sky background. \\
The spectra were reduced using the dedicated GIRAFFE ESO pipeline\footnote{\url{https://www.eso.org/sci/software/pipelines/giraffe/giraffe-pipe-recipes.html}} that performs bias subtraction, flat-fielding, wavelength calibration, and spectral extraction. For each exposure, the spectra of sky regions were median-combined together, and the derived master-sky spectrum was subtracted from each stellar spectrum. \\
Atmospheric parameters (effective temperature, surface gravity, and microturbolent velocity) for all sample
stars were adopted from \citetalias{alvarez_22}, but for the convenience of the reader here we summarize the procedure we adopted. We used photometric information from Gaia early Data Release 3 (\citealt{Gaia_16}, \citeyear{Gaia_21}). The spatial distribution of the sample relative to the cluster center is illustrated in Figure \ref{fig:distrib_rad}.\\
Effective temperatures ($T_{\mathrm{eff}}$) were computed using the empirical $(BP-RP)_0-T_{\mathrm{eff}}$ relation by \citet{mucc_21}, based on the InfraRed Flux Method, and assuming a color excess of $E(B-V)=0.12\pm0.02$ (\citealt{harris10}). In the calculation of the dereddened color $(BP-RP)_0$ we followed the scheme proposed by \citet{Gaia_18}. Internal errors in $T_{\mathrm{eff}}$ due to the uncertainties in photometric data, reddening and $(BP-RP)_0-T_{\mathrm{eff}}$ relation are in the range 85$-$115 K.\\
Surface gravities ($\log g$) were estimated from the Stefan-Boltzmann relation, adopting a typical mass of $0.80\;\mathrm{M_{\odot}}$, assuming the photometric $T_{\mathrm{eff}}$, the bolometric corrections for the dereddened G-band magnitude from \citet{andrae_18}, and a true distance modulus DM$_0=13.70\pm0.06$ (\citealt{principe06}). We computed the uncertainties in gravities by propagating the uncertainties in $T_{\mathrm{eff}}$, distance modulus and photometry. These uncertainties are of the order of $0.1$ dex. We would like to point out that an incorrect attribution of the targets to an evolutionary stage (AGB stars attributed to RGB sequence) has a negligible impact on the derived abundances: indeed a difference of 0.2 $M_{\odot}$ in the attribution mass leads to a modification in $\log g$ of $\sim$ 0.1, corresponding to a variation in the measured Mg, Al, Si, and Fe abundances of about 0.005 dex or less. To assess the impact of 0.1 dex change in $\log g$ on elemental abundances, we performed calculations while keeping the other atmospheric parameters fixed to their best values and only varying $\log g$ by the specified value. Our results indicate that such a difference leads to an extremely small change in the measured abundance ratios, amounting to less than 0.005 dex. This negligible variation arises from the fact that all the measured elements (Fe, Mg, Al, and Si) are in their neutral stage, rendering them almost insensitive to variations in $\log g$. \\
Microturbolent velocities ($v_t$) were obtained adopting the $\log g$-$v_t$ calibration by \citet{kirby_09}. This relation provides values of $v_t$ of about 1.6-2.0 km s$^{-1}$. We assumed a conservative error of $0.2$ km s$^{-1}$ in the determination of $v_t$ uncertainties.  All the relevant information about the observed targets (Id, Gaia G magnitude, radial velocity, and the derived atmospheric parameters) are reported in Table 1 in \citetalias{alvarez_22}.

\section{Abundance analysis} \label{sec:analysis}
In this work we adopted Fe and Mg abundances from \citetalias{alvarez_22}, while we derived abundances for Al and Si. In Table $\ref{table:2}$ are reported all the obtained elemental abundances.\\
Chemical analysis was performed using one-dimensional, Local Thermodynamic Equilibrium (LTE), plane-parallel geometry model atmospheres computed with the code ATLAS9 (\citealt{castelli04}) that treats the line opacity through the opacity distribution functions (ODF) method. All the models are calculated using the ODFs computed by \citet{castelli04} with $\alpha$-enhanced chemical composition and without the inclusion of the approximate overshooting in the calculation of the convective flux. \\
Si abundances were derived through the comparison between measured and theoretical equivalent widths (EWs) using the code GALA (\citealt{mucc_13}). We measured the EWs of selected lines with the code \texttt{DAOSPEC} (\citealt{stetson_08}) through the wrapper \texttt{4DAO} (\citealt{4dao}). Our lines were selected in order to be unblended and not saturated at the resolution of the GIRAFFE setups. The atomic data for our transitions are from the Kurucz/Castelli linelist \footnote{\url{https://wwwuser.oats.inaf.it/castelli/linelists.html}}. \\
Al abundances were derived using our own code \texttt{SALVADOR}, which performs a $\chi^2$ minimization between the observed line and a grid of suitable synthetic spectra calculated on the fly using the code \texttt{SYNTHE} (\citealt{kurucz_05}) in which only the Al abundance is varying. Al abundances were derived through spectral synthesis and not via EW, as we did for Si, because the Al doublet at 7835-7836 $\AA$ is contaminated by CN lines. At low metallicities, the impact of CN contamination is negligible; however, as metallicity increases, its impact becomes more pronounced\footnote{We would like to remark that the CN contamination does not affect the Fe, Mg, and Si at any metallicity.}. Since most of the stars in our sample do not have published C and N abundances, we fixed [N/Fe] $= +1.5$ dex as reasonable N value (according to \citealt{marino_12}) and treated C as free parameter to fit the CN affecting the Al doublet. Taking into account these assumptions, [C/Fe] abundance ratios between $-0.5$ and $+0.3$ dex provide the best fits to the CN lines.\\
Finally, all the derived abundance ratios are referred to the solar abundances of \citet{grev_98}.\\
We again followed the same approach used in \citetalias{alvarez_22} to estimate star-to-star uncertainties associated to the chemical abundances.\\
Internal errors, associated to the measurement process, were estimated as the line-to-line scatter divided by the root mean square of the number of lines. For Si, when only one line was available, we calculated the internal error by varying the EW of our lines of $1\sigma_{EW}$ (i.e., the EW error provided by \texttt{DAOSPEC}). For Al we adopted $\sigma/\sqrt{2}$ for all the stars in which we used both lines, while for the stars in which only one line was available we estimated the internal error by resorting to Monte Carlo simulation. \\
Errors associated to the adopted atmospheric parameters were computed by recalculating chemical abundances varying only one parameter at a time by its uncertainty, and keeping the other parameters fixed to their best value. \\
The uncertainties of the abundance ratios [Al/Fe] and [Si/Fe] were obtained following the Equation 2 in \citetalias{alvarez_22}.

\begin{table*}
\caption{Abundance ratios for the GIRAFFE targets of $\omega$ Centauri.}
\label{table:2}      
\centering          
\begin{tabular}{c  c  c  c  c}
\hline\hline \\ [-1.5ex]       
ID & [Fe/H] & [Mg/Fe] & [Al/Fe] & [Si/Fe]  \\
Sun & 7.50 & 7.58 & 6.47 & 7.55 \\ [0.5ex] 
\hline \\ [-1.5ex]
48$\_$NDC & $-1.92\pm 0.07$ & $0.42\pm 0.04$ & $0.51\pm 0.05$ & $0.29\pm 0.10$ \\ [0.4ex] 
74$\_$NDC & $-1.93\pm 0.10$ & $0.44\pm 0.03$ & $0.57\pm 0.06$ & $0.41\pm 0.10$ \\ [0.4ex]
84$\_$NDC & $-1.57\pm 0.07$ & - & $0.29\pm 0.07$ & $0.48\pm 0.10$ \\ [0.4ex]
161$\_$NDC & $-1.74\pm 0.10$ & $0.42\pm 0.02$ & $0.05\pm 0.07$ & $0.26\pm 0.09$ \\ [0.4ex]
182$\_$NDC & $-1.53\pm 0.07$ & $0.51\pm 0.05$ & $0.41\pm 0.04$ & $0.34\pm 0.08$ \\ [0.4ex]
357$\_$NDC & $-0.97\pm 0.06$ & - & $1.21\pm 0.06$ & $0.41\pm 0.12$ \\ [0.4ex]
480$\_$NDC & $-1.02\pm 0.11$ & - & $1.46\pm 0.06$ & $0.44\pm 0.14$ \\ [0.4ex]
27048$\_$J10 & $-1.59\pm 0.10$ & $-0.07\pm 0.05$ & $1.35\pm 0.05$ & $0.57\pm 0.09$ \\ [0.4ex] 
27094$\_$J10 & $-1.83\pm 0.11$ & $0.44\pm 0.06$ & $0.43\pm 0.08$ & $0.50\pm 0.10$ \\ [0.4ex]
29085$\_$J10 & $-1.75\pm 0.10$ & $-0.37\pm 0.07$ & $1.16\pm 0.06$ & $0.55\pm 0.10$ \\ [0.4ex]
\hline                  
\end{tabular}
\tablefoot{This is a portion of the table, available in its entirety in the electronic form.}
\end{table*}

\section{Results} \label{results}
Depending on the metallicity range, $\omega$ Centauri exhibits (anti-) correlations with different amplitudes. Given the wide range of metallicities present in the system, it is crucial to analyze the chemical anomalies not only as a whole, but also in distinct metallicity regimes. This approach will provide a more comprehensive understanding of the mechanisms underlying the complex chemical patterns observed in the $\omega$ Centauri.

\subsection{Fe, Mg, Al, and Si abundances}
\label{sec:ferro}
\begin{figure*}
   \centering
   \includegraphics[width=12 cm]{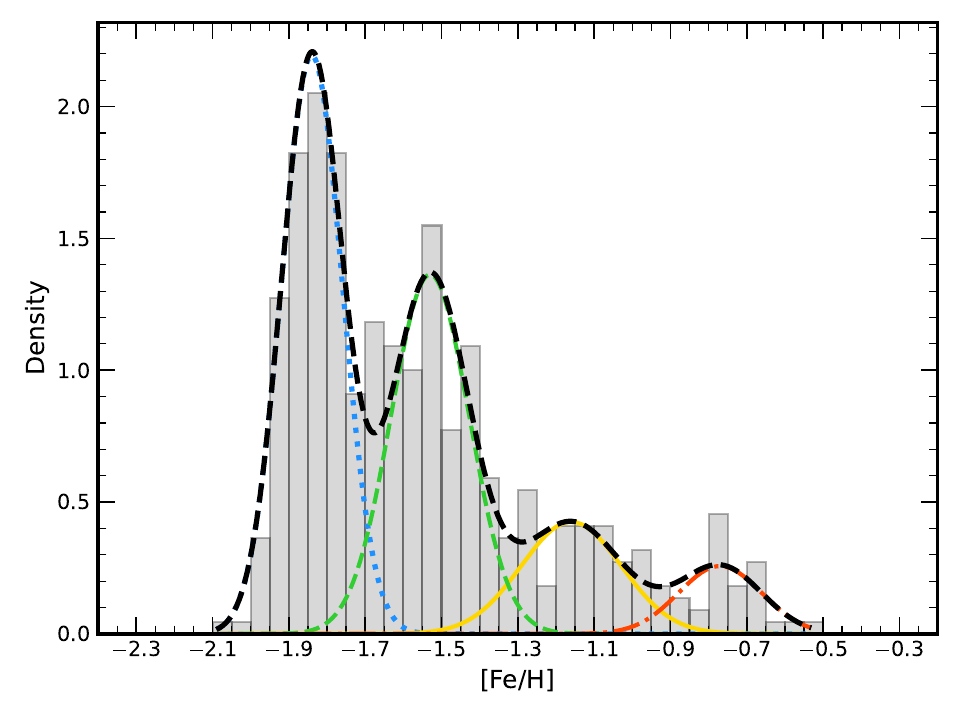}
      \caption{The histogram displays the MDF of $\omega$ Centauri. Also shown as dashed black line the Gaussian kernel fit of the distribution. Four Gaussian components can be identified. They represent the MP, M-int1, M-int2, and MR sub-populations, and are plotted in the figure as dotted, dashed, solid, and dashdotted lines respectively.}
         \label{fig:Hist_Fe}
\end{figure*}

According to the literature, $\omega$ Centauri hosts stars covering a broad range of metallicities. The metallicity distribution function (MDF hereafter) that we found in our investigation is shown in Figure $\ref{fig:Hist_Fe}$. In particular, to identify distinct populations in the data, we employed the \texttt{scikit-learn}\footnote{\url{https://scikit-learn.org/stable/modules/mixture.html}} implementation of Gaussian Mixture Models (GMM), which allowed us to identify four distinct groups corresponding to the following peaks in [Fe/H]: $-1.85, -1.55, -1.15$, and $-0.80$ dex. For comparison we considered the analyses performed by \citet{johnson_10} and \citet{meszaros_21} in which they studied a total sample of 855 and 1141 stars respectively, covering cluster's entire metallicity range. In particular, \citet{johnson_10} found the presence of five distinct metallicity peaks that are located at [Fe/H] = $-1.75, -1.50, -1.15, -1.05$, and $-0.75$ dex (in their distribution the peaks at $-1.15$ and $-1.05$ dex were combined due to the difficult to separate the two populations), while \citet{meszaros_21} found four peaks at [Fe/H] = $-1.65, -1.35, -1.05$, and $-0.7$ dex. The three MDFs exhibit a good agreement with each other, well in terms of [Fe/H] extension and relative position and intensity of the peaks.\\
With the exception of the peak at $-1.15$ dex, our measurements of the metallicity peaks are slightly lower (by about 0.05-0.1 dex) compared to those reported by \citet{johnson_10}. On the other hand, the metallicity peaks in our study are systematically lower by about 0.1-0.2 dex than the ones from \citet{meszaros_21}. These discrepancies are likely to be attributed to the very different method and set of lines used to calculate the metallicities from the H band (\citealt{meszaros_21}) and optical spectra. \\
Finally, following the results obtained from our distribution and a nomenclature similar to the one adopted by \citet{sollima_05} we divided our population in the following four sub-populations: metal-poor (MP, [Fe/H]$\leqslant -1.69$ dex), metal-int1 (M-int1, $-1.68\leqslant$ [Fe/H] $\leqslant -1.34$ dex), metal-int2 (M-int2, $-1.33\leqslant$ [Fe/H] $\leqslant -0.94$ dex), and metal-rich (MR, [Fe/H]$\geqslant -0.93$ dex). In Table $\ref{tab:met}$ are reported the main information for the four sub-populations. 

\begin{table}
      \caption[]{The table lists the mean metallicity $\langle \textrm{[Fe/H]} \rangle$, its associated dispersion ($\sigma$), the sample size (N) and number fraction of the four metallicity sub-populations identified in $\omega$ Centauri (see text for details).}
         \label{tab:met}
        \centering          
\begin{tabular}{ccccc}
\hline\hline \\ [-1.5ex]   
            Group &  $\langle \textrm{[Fe/H]} \rangle$ & $\sigma$ & N & fraction \\
             &  (dex) & (dex) & number &  \\
            \hline
            Metal-poor (MP)& -1.85 & 0.08 & 193 & 0.44     \\
            Metal-int1 (M-int1)& -1.55 & 0.10 & 153 & 0.35     \\
            Metal-int2 (M-int2) & -1.15 & 0.13 & 63 & 0.14     \\
            Metal-rich (MR) & -0.80 & 0.11 & 30 & 0.07     \\
\hline                  
\end{tabular}
\end{table}

\begin{figure*}
   \centering
   \includegraphics[width=\textwidth]{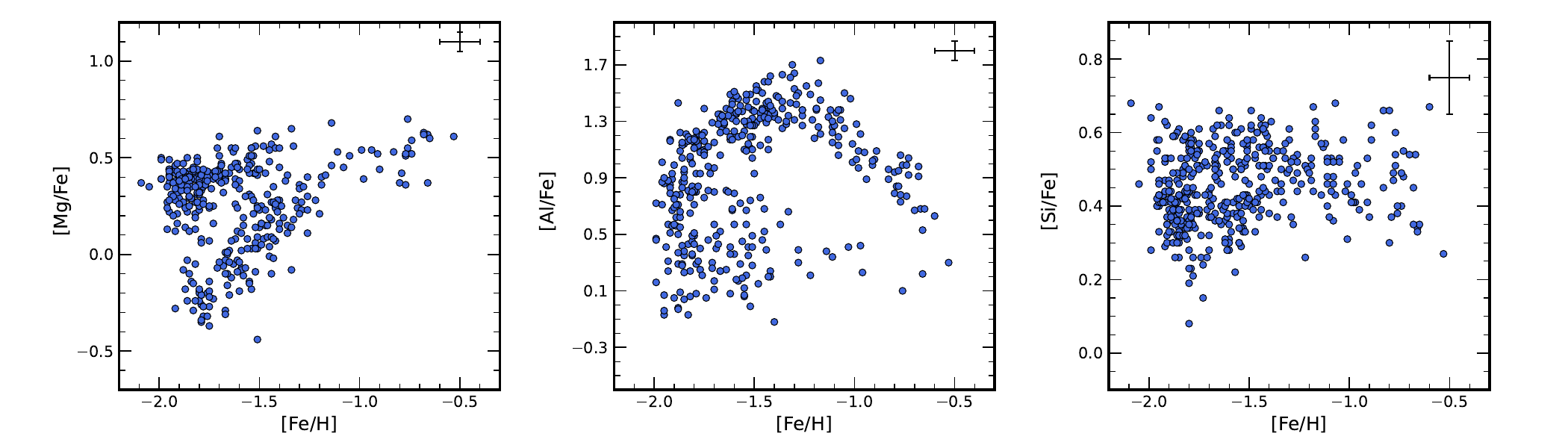}
      \caption{Distribution of [Mg/Fe] (left panel), [Al/Fe] (middle panel), and [Si/Fe] (right panel) as a function of [Fe/H] for the stars belonging to our sample. The error bar in the top right corner represents the typical error associated to the measurements.}
         \label{fig:3panel}
\end{figure*}

In Figure $\ref{fig:3panel}$ we can observe the behavior of [Mg/Fe], [Al/Fe], and [Si/Fe] as a function of [Fe/H] for the stars that we analyzed in this work. In the left panel, we can see that [Mg/Fe] distribution is split in two different branches, with the upper branch that covers a [Fe/H] range from $\sim -2.1$ dex up to $\sim -1.3$ dex and is characterized by enriched values of [Mg/Fe]. On the other hand, the lower branch covers a range of [Fe/H] from $\sim -1.9$ dex up to $\sim -0.5$ dex, with Mg abundances ranging from sub-solar values up to the highest values [Mg/Fe] $\sim +0.6$ dex at the highest metallicities. For the [Al/Fe] distribution the behavior is completely different. Indeed, in the MP sub-population we have a large spread in [Al/Fe] with abundances from [Al/Fe] $\sim -0.15$ dex up to [Al/Fe] $\sim +1.3$ dex. At higher metallicities ([Fe/H] $>-1.7$ dex) there is the presence of a branch that reaches its maximum extension in the Al abundance at [Fe/H] $\sim -1.3$ dex and then there is a clear decrease in the Al abundances down to [Al/Fe] $\sim +0.6$ in the MR sub-population. Finally, we have a minor group of stars with [Al/Fe] $<0.9$ dex in the M-int1, M-int2, and MR sub-populations. The number of stars in this latter group diminish significantly at the highest metallicities, with only three stars present in the MR sub-population. Finally, the behavior of [Si/Fe] as a function of [Fe/H] is bimodal at [Fe/H]$< -1.3$ dex, even though the separation between the two branches is not so evident (they are separated by $\sim 0.2$ dex). At higher metallicities all the stars are characterized by enhanced [Si/Fe] with a spread fully compatible with the typical uncertainties. 

\subsection{Mg-Al anti-correlation}
\begin{figure*}[ht!]
   \centering
   \includegraphics[width=14 cm]{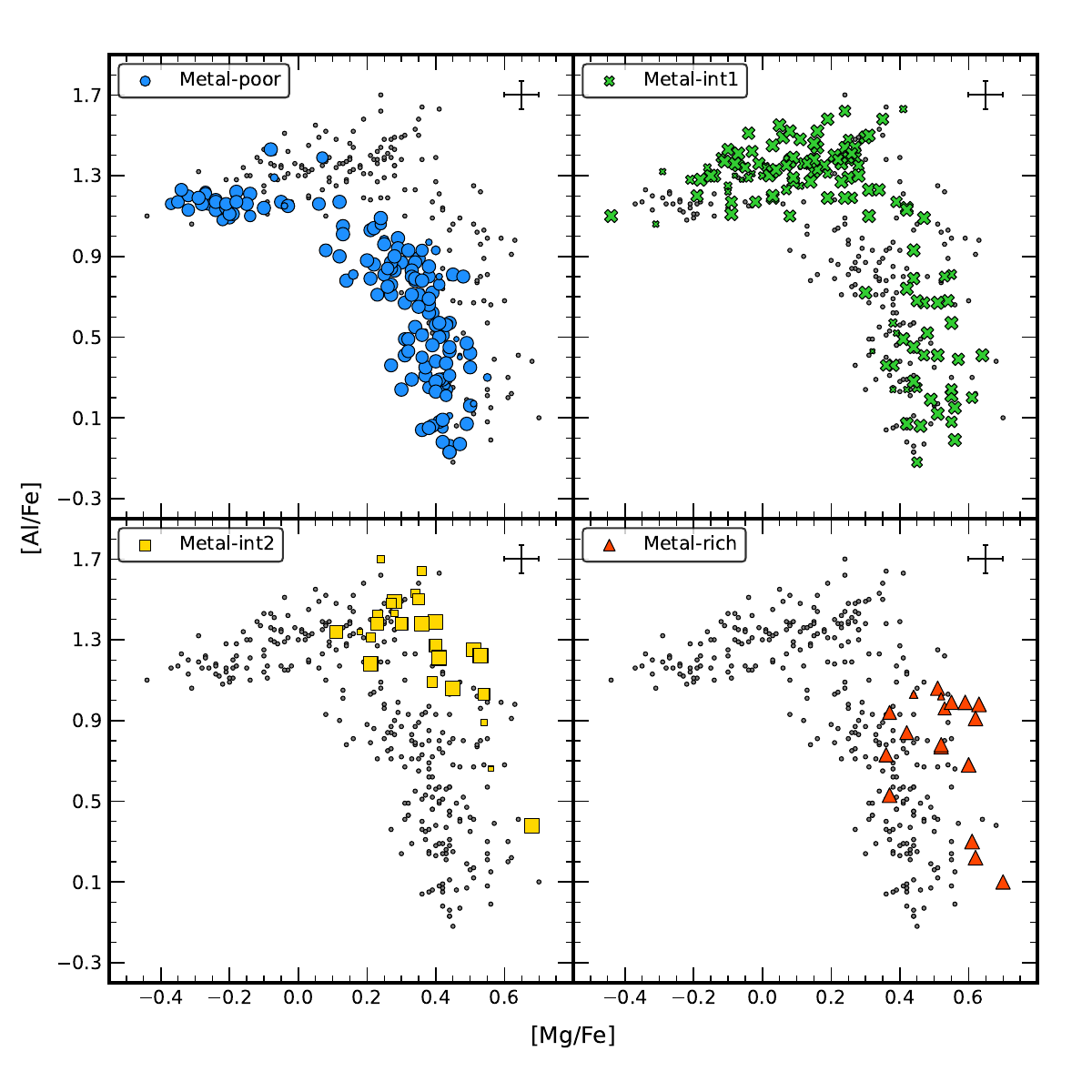}
      \caption{The four panels shown depict the trend of [Mg/Fe] as a function of [Al/Fe] for the MP, M-int1, M-int2, and MR sub-populations, respectively (from top to bottom, left to right). The size of each point indicates its probability of belonging to that particular metallicity sub-population - the larger the symbol, the higher the probability - while gray dots represent the entire sample.  The error bar in the top right corner represents the typical measurement error associated with the data.}
         \label{fig:Mg_Al}
\end{figure*}

We observe a large spread in both [Mg/Fe] and [Al/Fe] abundances ratios, with [Mg/Fe] ranging from 0.70 dex down to subsolar values (the minimum abundance value is $-0.44$ dex) with a mean value [Mg/Fe] $= +0.26$ dex ($\sigma = 0.23$ dex), while [Al/Fe] ranges from  $+1.70$ dex down to $-0.15$ dex with a mean value [Al/Fe] $=+0.93$ dex ($\sigma = 0.44$ dex). For the 323 stars for which both Mg and Al abundances are available a discrete Mg-Al anti-correlation can be detected, as can be seen in Figure $\ref{fig:Mg_Al}$. In particular, groups of stars with different metallicity exhibit different Mg-Al distributions. Especially, the MP and M-int1 sub-populations show a clear Mg-Al anti-correlation (see top panels of Figure \ref{fig:Mg_Al}), whereas the other two sub-populations show a chemical anomaly that is either less clear or not present at all (see bottom panels of Figure \ref{fig:Mg_Al}).

{\bf 1.} In the 144 stars that make up the MP sub-population, we recognize the presence of a distinct Mg-Al anticorrelation, with all of the Mg-poor ([Mg/Fe] $< 0.0$ dex) stars having [Al/Fe] $\sim +1.15$ dex. On the other hand, the Mg-rich stars ([Mg/Fe]$> 0.0$ dex) are distributed from [Mg/Fe] $\sim 0.15$ up to [Mg/Fe] $\sim +0.5$ dex and they cover a wide range of [Al/Fe] (from $\sim -0.15$ up to $\sim +1$ dex). Finally, we observe that the two groups of stars are clearly separated by a sort of gap in [Mg/Fe] between $-0.1$ and $+0.1$ dex. For the Mg-rich group (112 out of 144 stars) we ran a Spearman correlation test and calculated the correlation coefficient ($C_s$) and the corresponding two-tailed probability that an absolute value $C_s$ larger than the observed one can be obtained starting from uncorrelated variables, in order to better quantify the amplitude of this anti-correlation. In this instance, we found $C_s = -0.70$, which resulted in a zero chance that the observed anti-correlation could have come from uncorrelated data.\\
    \begin{figure}[h]
     \centering
    \includegraphics[width=9 cm]{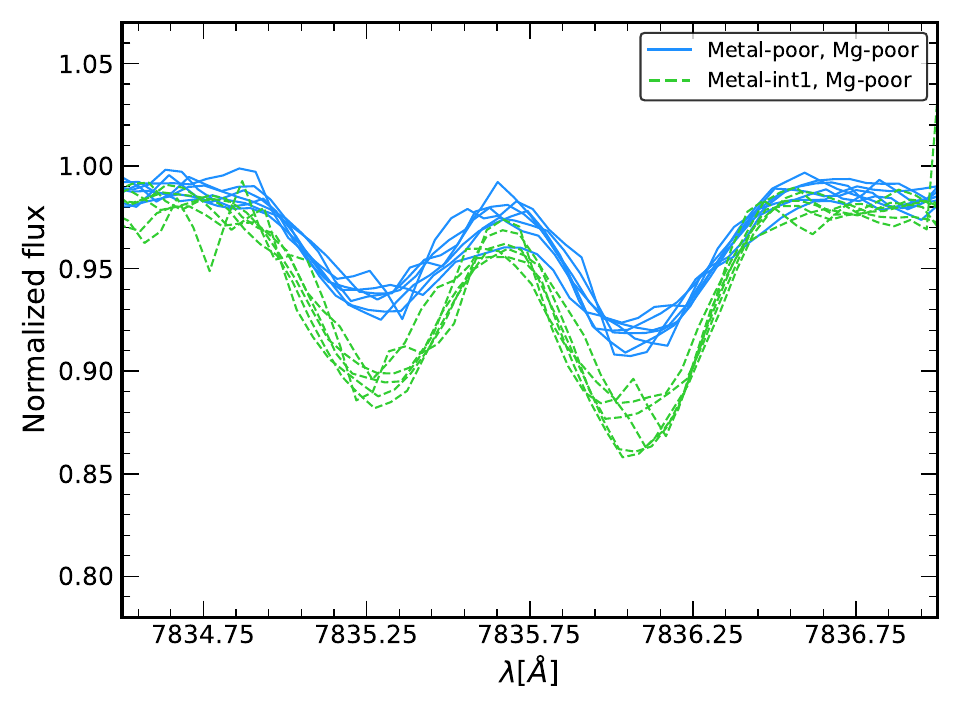}
          \caption{The figure shows the spectral region around the Al doublet at 7835-7836 $\AA$ for sample stars with similar atmospheric parameters. The solid blue lines denote stars belonging to the Mg-poor group in the MP sub-population, whereas the dashed green lines represent again Mg-poor stars but in the M-int1 sub-population. A clear difference in the strength of the Al lines can be appreciated between the two metallicity sub-groups.}
             \label{fig:Al_lines}
    \end{figure}
{\bf 2.} We can clearly discern a Mg-Al anti-correlation also in the M-int1 sub-population (135 stars). We can easily distinguish two principal groups of stars in this metallicity class, which have similarities with the previous group. The first sub-group has high [Mg/Fe] values (ranging from $\sim+0.4$ to $\sim+0.6$ dex) and a wide range of [Al/Fe] values (from $\sim+0.3$ to $\sim+1.0$ dex). The second group, however, has a smaller range of [Al/Fe] (from $\sim+1.3$ up to $\sim+1.7$ dex) but a wider range of [Mg/Fe] (from $\sim+0.35$ down to $\sim-0.2$ dex). There are a considerable number of Mg-poor stars in this latter group. The Spearman test was also run in this instance, and the results showed that $C_s = -0.60$ and a p-value that was consistent with zero.\\
From the top panels of Figure \ref{fig:Mg_Al} we can clearly note that all the stars in the Mg-poor group of MP sub-population have lower Al abundances ([Al/Fe] $\sim 0.2$ dex lower) than the stars in the same Mg-poor group belonging to the M-int1 sub-population. We investigated whether this effect is artificial or not, but neither the stellar parameters nor the evolutionary state were found to be related. Additionally, we can see the spectra of some stars from the MP and M-int1 sub-populations in Figure $\ref{fig:Al_lines}$ around the Al doublet. Even though they are members of different sub-populations, all stars share similar atmospheric parameters and metallicity. In particular, stars were selected to have metallicities between $-1.78\leqslant$ [Fe/H] $\leqslant -1.69$ dex and $-1.68 \leqslant$ [Fe/H] $\leqslant -1.63$ dex for the MP and the M-int1 sub-populations, respectively. The figure doubtless demonstrates that Al lines are weaker in the Mg-poor stars corresponding to the MP sub-population.\\
{\bf 3.} The existence of the Mg-Al anti-correlation in the 26 stars that make up the M-int2 sub-population is less certain. The abundances of the stars range from $\sim+0.1$ to $\sim+0.6$ dex for [Mg/Fe] and from $\sim+1$ to $\sim+1.7$ dex for [Al/Fe], with Mg abundances that are between those of the Mg-rich stars in the previous two sub-populations. One star only stands out among these values, having much lower Al and greater Mg abundances than the other stars in the same group. We are not able to tell whether discrete star groups are present in this sub-population. \\
{\bf 4.} Finally, for the MR sub-population (18 stars) there is no evidence of an anti-correlation, with the stars that have [Mg/Fe] $> \sim +0.35$ dex and [Al/Fe] $< \sim +1.1$ dex. This result is not surprising, since the presence of Mg-Al anti-correlation is detected in almost all GCs that are more metal-poor than [Fe/H] $\lesssim -1$  (\citealt{shetrone_96}; \citealt{meszaros15}; \citealt{pancino17}).
In summary, an anti-correlation between the abundance of Mg and Al has been observed in $\omega$ Centauri, with the strength of the correlation being dependent on the metallicity of the individual stars being considered.
A clear anti-correlation can be detected, in particular, for all stars with [Fe/H] $\lesssim -1.3$ dex. However, for stars in the range between $-1.3 <$ [Fe/H] $< -0.9$ dex, the anti-correlation is less obvious, and ultimately, for stars more metal-rich than [Fe/H] $\gtrsim -0.9$ dex, there is no sign of the Mg-Al anti-correlation.\\
Different studies made by \citet{NDC95}, \citet{smith_00} and \citet{meszaros_21} found evidence of a significant spread in the Mg and Al abundances in $\omega$ Centauri. In particular, \citet{meszaros_21} found evidence of an extended Mg-Al anti-correlation among the stars with [Fe/H] $< -1.2$ dex. However, our study represents the first clear detection of a discrete Mg-Al anti-correlation, specifically within the MP and M-int1 sub-populations, with different Mg and Al distributions at different metallicities. Furthermore, 73 $\%$ of our stars have [Al/Fe] $> 0.5$ dex. The existence of a such fraction of stars with high values of Al indicates that the majority of stars in our sample belong to the so-called 2P, that was born from material processed in the polluter stars by the MgAl chain at high temperatures (\citealt{ventura_16}; \citealt{dellagli_18}).

\begin{figure*}[!h]
   \centering
   \includegraphics[width=14 cm]{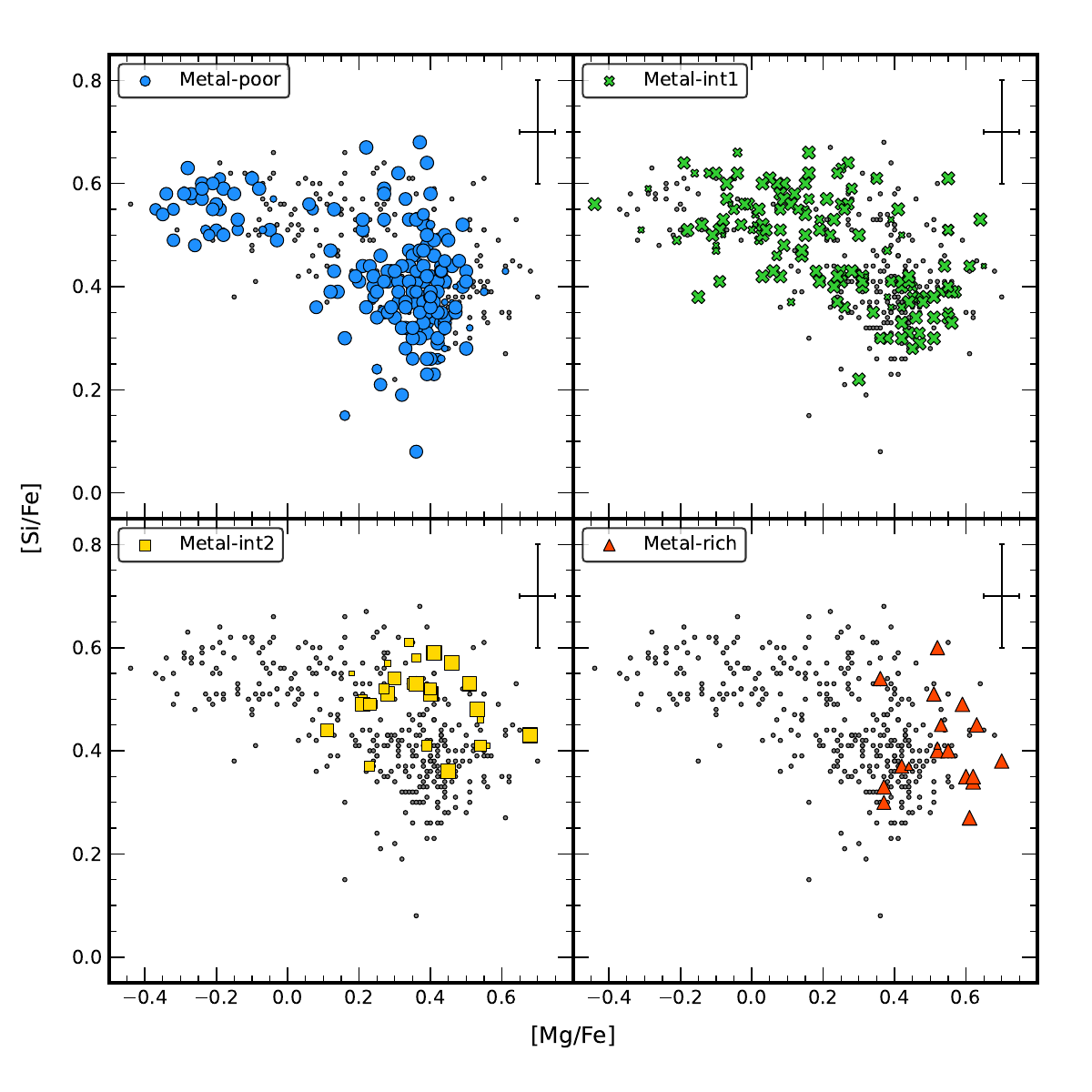}
      \caption{As for Figure \ref{fig:Mg_Al}, but for [Si/Fe] and [Al/Fe].}
         \label{fig:Mg_Si}
\end{figure*}

\subsection{Mg-Si anti-correlation and Al-Si correlation}
\begin{figure*}
   \centering
   \includegraphics[width=14 cm]{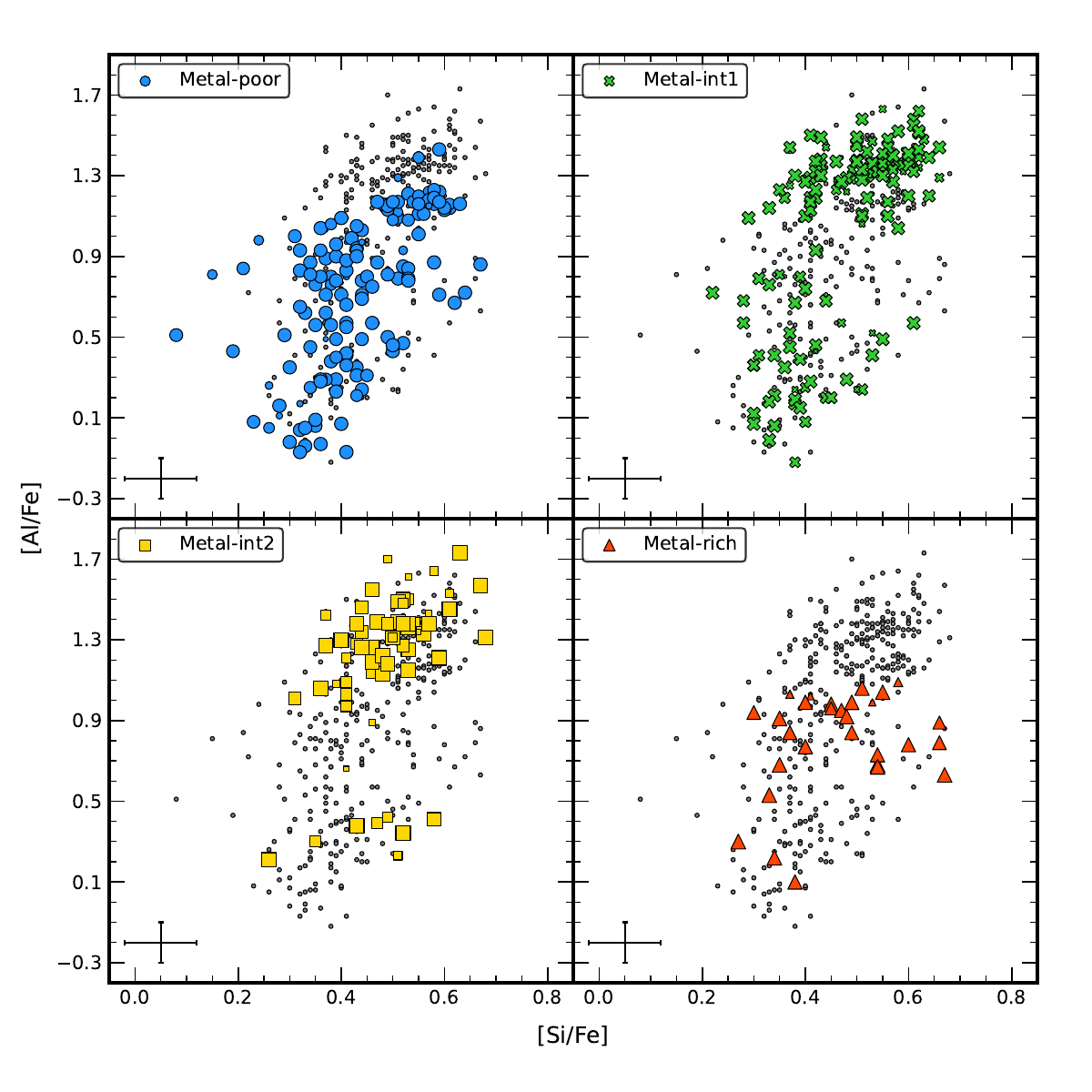}
      \caption{As for Figure \ref{fig:Mg_Al}, but for [Mg/Fe] and [Si/Fe].}
         \label{fig:Al_Si}
\end{figure*}

For a total of 370 stars, the abundances of Mg and Si are simultaneously measured. With a mean value of +0.45 dex ($\sigma =  0.10$ dex), [Si/Fe] varies from $+0.68$ dex down to $+0.08$ dex. A Mg-Si anti-correlation is present in this sample, which is mostly contributed by the population's MP subgroup (see Figure $\ref{fig:Mg_Si}$). In fact, the Mg-poor and the Mg-rich stars may be distinguished clearly from one to another in this sub-population. The first group is distinguished by a mean value of [Si/Fe] $= +0.55$ dex, whereas the second group has a mean value of [Si/Fe] $+0.35$ dex, even if there are certain stars (the minority) with [Si/Fe] $> +0.5$ dex. In comparison to the Mg-rich group, the Mg-poor group is therefore increased by roughly $+0.2$ dex, which is significantly larger than the typical error associated to the [Si/Fe] measurements (0.1 dex). \\
It is worth noting that the Si-enhancement (Al-depletion) in $\omega$ Centauri is primarily observed in stars with low Mg abundances, particularly in the most metal-poor population. \\
We have used the Spearman correlation test to examine the Mg-Si anti-correlation in the MP sub-population, much like we did for the Mg-Al anti-correlation. Our results show that $C_s = 0.45$ and the p-value consistent with zero. This suggests that the presence of a Mg-Si anti-correlation in the MP sub-population of $\omega$ Centauri is real.\\
Besides the MP sub-population, only the M-int1 sub-population displays a clear anti-correlation between Mg and Si abundances, despite the challenge of distinguishing between the various sub-groups within this population.\\
An Al-Si correlation can be observed in $\omega$ Centauri, as we can appreciate in Figure $\ref{fig:Al_Si}$. In this case, Si and Al are available at the same time for a total of 381 stars.\\
Numerous previous studies have theorized that the Al-Si correlation in GCs is the result of a leakage from the MgAl chain into $^{28}\textrm{Si}$ via proton capture reaction at extremely high temperatures (e.g., \citealt{yong_05}; \citealt{meszaros15}; \citealt{masseron_19}). \\
All of the (anti-) correlations associated with the MgAl chain are seen in the MP and M-int1 sub-populations of $\omega$ Centauri, when Mg, Al, and Si are analyzed together while taking into account the various metallicity groups. Particularly, the Mg-depleted and Al (mildly) enhanced stars, which are assumed to be the product of the extreme MgAl processing in the polluter stars, correspond to the bulk of Si-enhanced stars. The proton capture processes in these stars took place at temperatures greater than 10$^8$ K. Contrarily, only a (weak) Mg-Al anti-correlation is found in the M-int2 sub-population, whereas the other two anti-correlations are either nonexistent or just weakly confirmed.
Finally, there is no indication of any chemical anomaly linked to the MgAl chain in the MR sub-population. This result is expected given that the polluter stars are unable to attain the temperatures required to begin the Mg destruction through proton capture at those metallicities ([Fe/H] $\gtrsim -0.9$ dex).

\section{Comparison with \citet{meszaros_21}} \label{comparison_meszaros}
\begin{figure*}
   \centering
   \includegraphics[width= 14 cm]{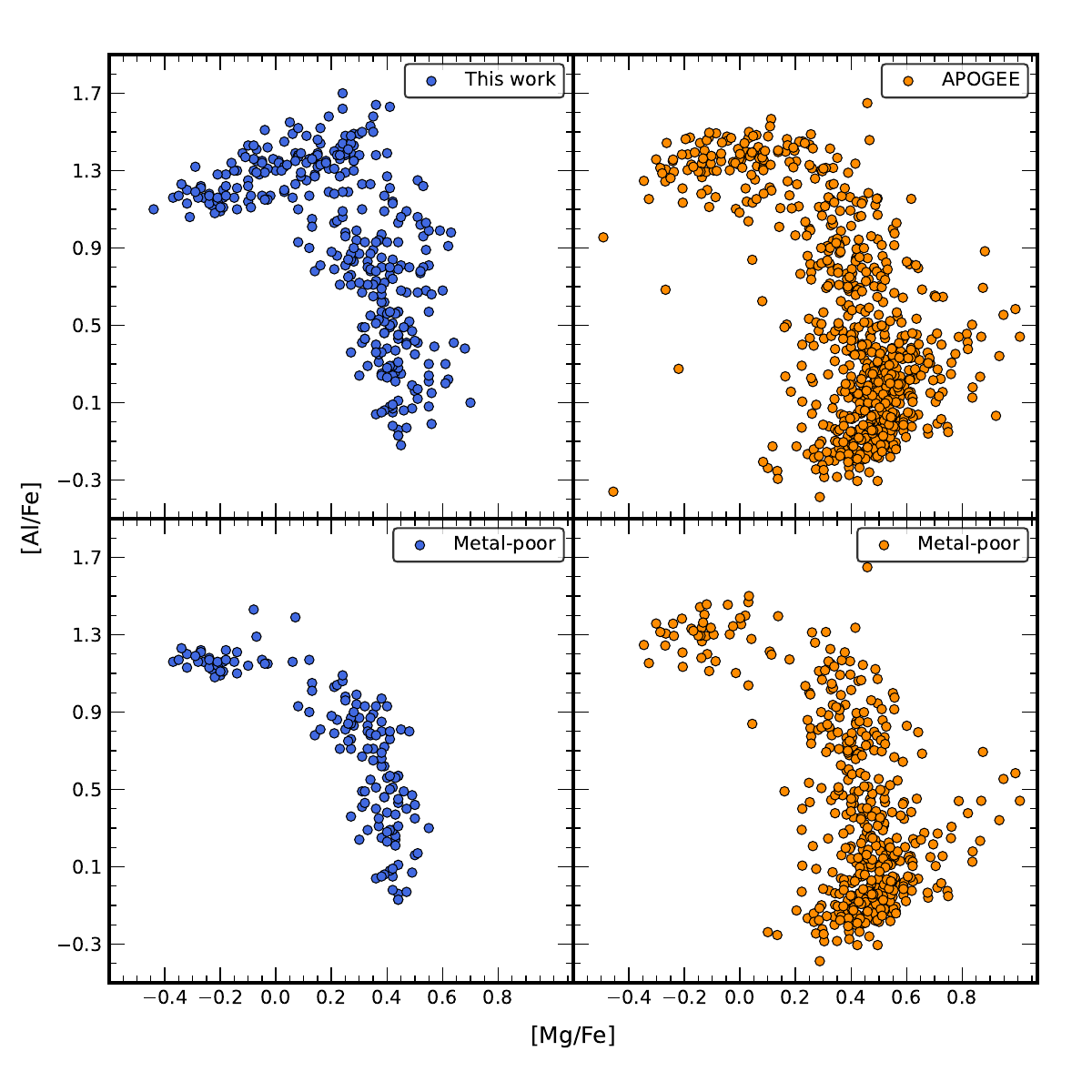}
      \caption{Trend of [Mg/Fe] as a function of [Al/Fe] for the stars analyzed in this study (top-left panel), and for the stars analyzed by \citet{meszaros_21} (top-right panel). In the bottom panels are displayed only the stars belonging to the MP sub-population for each sample, respectively.}
         \label{fig:4panel_MgAl}
\end{figure*}

\begin{figure}
   \centering
   \includegraphics[width= 9 cm]{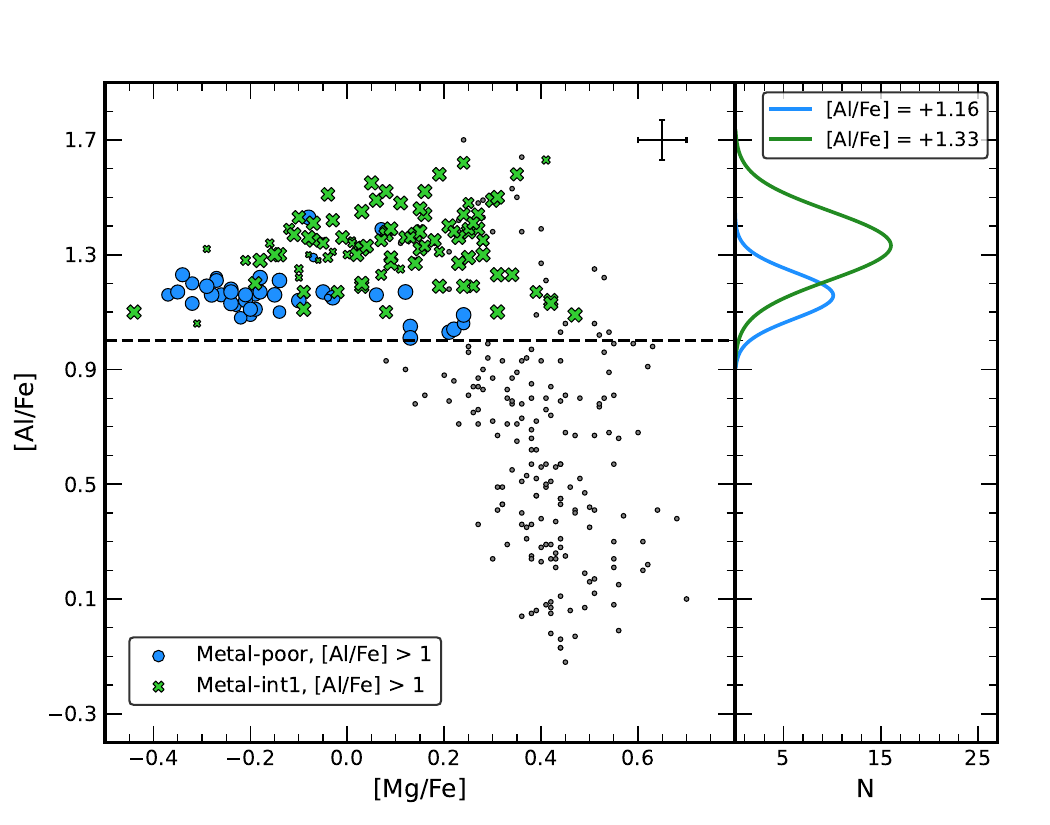}
      \caption{Trend of [Mg/Fe] as a function of [Al/Fe] for the stars analyzed in this study. The blue dots and the green crosses represent the MP and M-int1 stars with [Al/Fe] $>$ 1 dex, respectively.  The gray dots represent the entire sample. The error bar in the top right corner represents the
      typical measurement error associated with the data. The dashed line indicates the value [Al/Fe] = 1 dex. The distributions of the two sub-samples are shown in the right panel with the corresponding colors. }
         \label{fig:disc_MgAl}
\end{figure}

\begin{figure*}
   \centering
   \includegraphics[width=\textwidth]{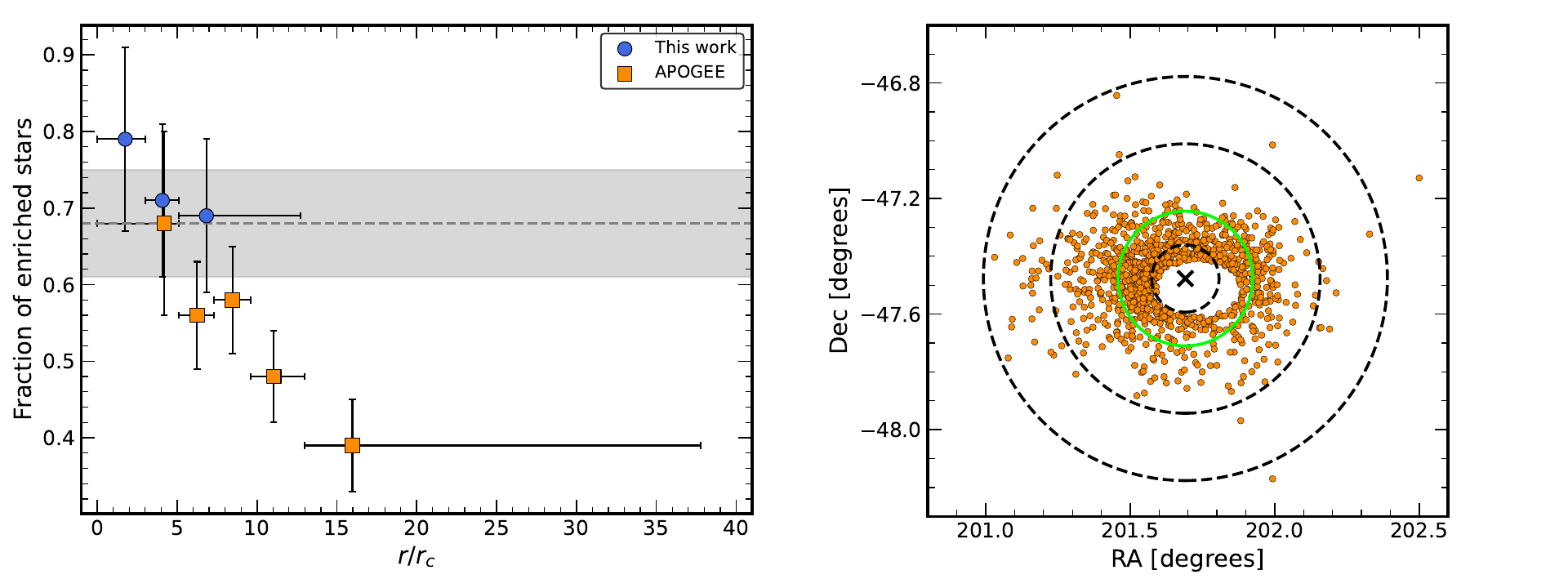}
   \caption{Left panel: fraction of the enriched stars as a function of the distance from the cluster center, as defined by \citet{vanleeuwen_00}. The blue circles represent the stars analyzed in this study, whereas the orange squares display the stars studied by \citet{meszaros_21}. In the Y axis, the error bars represent the error associated to the fraction of enriched stars, while in the X axis represent the extension of the stars contained in each radial ring. The dashed gray line and the gray area show the mean and the standard deviation for the genuine GCs observed so far (\citealt{bastian_15}).\\
   Right panel: Coordinate positions of stars analyzed by \citet{meszaros_21}. The black cross represents the same cluster center used in Figure \ref{fig:distrib_rad}. The dashed black circles show 5, 20, and 30 times the core radius ($r_c = 1\rlap{.}^{'}40$; \citealt{harris_96}). The green circle represents 10 times the core radius and it encloses 554 out of 982 stars.}
   \label{fig:frac_enriched}
\end{figure*}

\begin{figure*}
   \centering
   \includegraphics[width=\textwidth]{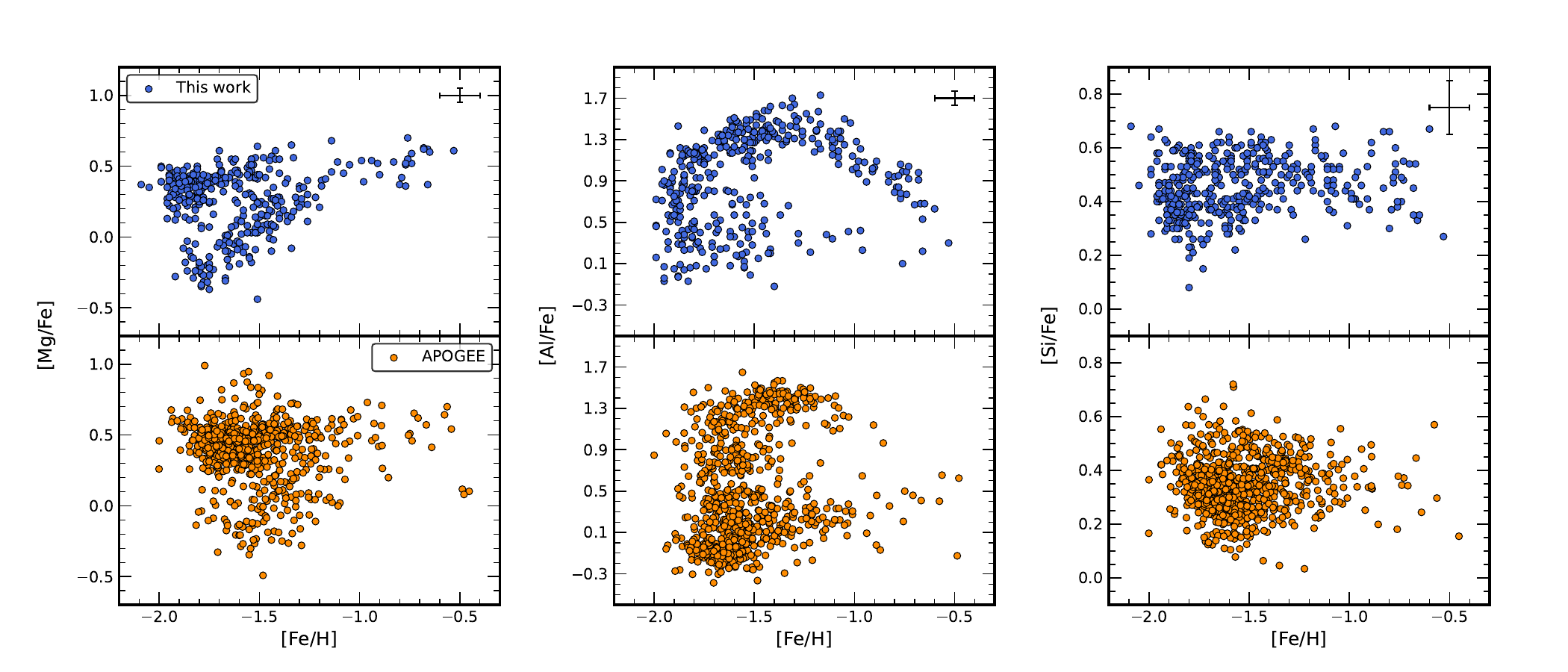}
      \caption{Distribution of [Mg/Fe] (left panel), [Al/Fe] (middle panel), and [Si/Fe] (right panel) as a function of [Fe/H]. In the top are displayed the stars here analyzed (blue dots), while in the bottom the stars analyzed by \citet{meszaros_21} (orange dots). The error bar in the top right corner in the top figures represents the typical error associated with the measurements.}
         \label{fig:6panel}
\end{figure*}

A large spectroscopic analysis measuring Mg, Al, and Si abundance variations in $\omega$ Centauri was performed by \citet{meszaros_21}. In particular, they studied a total sample of 982 stars with high signal-to-noise (S/N $> 70$), observed by the SDSS-IV/APOGEE-2 survey (\citealt{majewski_17}). \citet{meszaros_21} found behaviors of the three abundance ratios qualitatively similar to our ones. They found a Mg-Al anti-correlation, the shape of which clearly depends on the metallicity of the considered stars (see their Figure 4). In particular, at high metallicities ([Fe/H] $> -1.2$ dex) the presence of a Mg-Al anti-correlation is less evident, but with a bimodal distribution in the Al abundances; on the other hand, at the highest values of [Fe/H] the Mg-Al anti-correlation disappears and the Al abundances are nearly constant. These findings are well in agreement with our results and indicate a weakening of the Mg-Al anti-correlation extension towards higher metallicities. Despite the similarities in the morphology, some relevant differences between our results and those by \citet{meszaros_21} are present (see top panel of Figure \ref{fig:4panel_MgAl}). \\
{\bf -} Differences in the Mg-Al anti-correlation in the MP sub-population: if we consider the MP sub-population in both samples \footnote{Note that \citet{meszaros_21} define their MP sub-population at [Fe/H] $\leqslant -1.5$ dex.} (see bottom panel of Figure \ref{fig:4panel_MgAl}) we can observe some interesting differences. In both samples we can detect a clear gap between the Mg-poor and Mg-rich stars. However, at variance with us, in the sub-group of stars with [Mg/Fe] $> 0.0$ dex \citet{meszaros_21} do not detect a clear Mg-Al anti-correlation. Indeed they observe almost a constant [Mg/Fe], with a large spread in [Al/Fe] (from [Al/Fe]$\sim -0.25$ dex up to [Al/Fe] $\sim +1.25$ dex). \\
{\bf -} Discreteness of Al-rich group: if we consider the MP and M-int1 sub-populations, we can observe that Al-rich stars are well separated by a gap. To test whether this observed discreteness is real or not, in our sample we considered the most Al-rich stars ([Al/Fe] $> +1.$ dex) in both MP and M-int1 sub-populations, and we measured if they are compatible to have the same Al. We found a mean value [Al/Fe] $= +1.16 \pm 0.01$ dex ($\sigma = 0.08$ dex) for the MP sub-sample, and [Al/Fe] $= +1.33 \pm 0.01$ dex ($\sigma = 0.11$ dex) for the M-int1 sub-sample (see Figure \ref{fig:disc_MgAl}). These values indicate that these two sub-samples are compatible at a level of 8.54 $\sigma$, thus strongly suggesting that they do not display the same Al abundances. On the other hand, this discreteness is not observed in the dataset of \citet{meszaros_21}. This difference could be due to our higher precision in the abundance measure that allows us to distinguish different abundance substructures.\\
{\bf -} Differences in the 2P/1P ratio: the most striking difference is in the fraction of 2P stars sampled by the two studies, with us having 2/3 of stars belonging to 2P, while \citet{meszaros_21} have only 1/2. We obtained these fractions of 2P by performing a population analysis, on both Mg-Al anti-correlations, using GMM algorithm. We considered the overall distributions on the Mg-Al plane for both anti-correlations, without making any division among the metallicity sub-populations. The aim of this type of analysis was not to uncover the presence of distinct sub-populations within the two distributions, but rather to separate the populations between so-called enriched (with high Al abundances) and primordial (with low Al abundances) stars. In the case of our Mg-Al anti-correlation, the GMM algorithm revealed that the enriched stars constitute the $73 \%$ ($\sigma = 6 \%$) of the entire population, while for the \citet{meszaros_21} Mg-Al anti-correlation, the enriched population forms the $52 \%$ ($\sigma = 3 \%$) of the total sample. To further investigate this result, we repeated the GMM analysis at various distances from the cluster center, taking into account the fact that our sample is more radially concentrated relative to the stars analyzed in \citet{meszaros_21}. Specifically, our targets are located at a distance of about 15 core radii, while the stars studied in \citet{meszaros_21} extend up to 30 core radii (as shown in the right panel of Figure \ref{fig:frac_enriched}). \\
To investigate the radial distribution of the percentage of enriched stars,  we divided our sample into three radial annuli: stars located within 3 core radii (105 stars), stars between 3 and 5 core radii (110 stars), and stars beyond 5 core radii (108 stars). In the case of \citet{meszaros_21} we examined stars located within 5 core radii (80 stars), between 5 and 7.3 core radii (169 stars), between 7.3 and 9.6 core radii (171 stars), between 9.6 and 13.0 core radii (174 stars), and beyond 13.0 core radii (169 stars). In the left panel of Figure $\ref{fig:frac_enriched}$ we can observe the fraction of enriched stars in both samples. In particular, in the innermost region our value is slightly higher (but within the errors) compared to the mean value observed in other GCs (see \citealt{bastian_15} for a detailed discussion). On the other hand, except for the value within 5 core radii, the fraction of enriched stars found by \citet{meszaros_21} are constantly below the mean value observed in other GCs, even though in the overlapping regions the two distributions are consistent within the uncertainties. This may be due to the different radial distribution of the two samples, as it is well known that 2P stars are more centrally concentrated than 1P stars in $\omega$ Centauri (as well as in many other clusters; e.g., \citealt{bellini_09}; \citealt{bastian_15}).\\
We speculate that the difference in the fraction of enriched stars between \citet{meszaros_21} and our study may partially explain why the Mg-Al anti-correlation exhibits distinct shapes. Indeed, \citet{meszaros_21} analyzed more external regions of $\omega$ Centauri, and the higher fraction of 1P stars in their sample could potentially contribute to explain the observed differences in the Mg-Al anti-correlation shape.\\
{\bf -} Differences in the behavior with [Fe/H]: in Figure \ref{fig:6panel} we can observe a comparison between the stars here analyzed and the ones studied by \citet{meszaros_21} for the distributions of [Mg/Fe], [Al/Fe], and [Si/Fe] as a function of [Fe/H]. If on one hand, we have similar behaviors in all three elements, there are also some interesting differences. In the [Mg/Fe] vs [Fe/H] plane the presence of the two branches in our sample is clear, while in the case of \citet{meszaros_21} their presence is barely visible. This effect may be attributed to the predominance of Mg-rich stars in their sample, as well as potential limitations in the measurement accuracy that could prevent a clear separation of the two branches in their analysis. Regarding [Al/Fe], our distribution and the one found by \citet{meszaros_21} covers a similar range of abundances. However, in the case of \citet{meszaros_21} the absence of Al-rich stars at [Fe/H] $> -1$ dex does not allow to observe if either is present or not a trend in the [Al/Fe] distribution against the metallicity. Moreover, their sample presents a population with [Al/Fe] $< \sim +0.5$ dex that constitutes half of the entire sample (these stars represent the 1P stars) and is distributed at almost every metallicity, with a decrease towards the highest metallicities. Finally, the [Si/Fe] distribution in the case of \citet{meszaros_21} is characterized by a constant and enhanced [Si/Fe] abundance ratio over the entire range of metallicity, at variance with the bimodality we observe in our sample at [Fe/H] $< -1.3$ dex. We would like to note that for the stars analyzed by \citet{meszaros_21} we applied their same criteria, limiting our selection only to stars with S/N $> 70$, T$_{eff} < 5500$ K, and with errors in the single abundances $< 0.2$ dex. Furthermore, it is worth mentioning that few stars were observed with [Fe/H] $> -1$ dex in the [X/Fe] (where X represents Mg, Al, and Si) versus [Fe/H] diagrams. 

\section{Discussion and conclusions} \label{discussion}
In this work, we investigated the multiple populations of $\omega$ Centauri by evaluating the effects of the MgAl cycle in the stars of this system. We derived the Fe, Mg, Al, and Si abundances for a total of 439 giant stars from the analysis of high-resolution spectra obtained with the multi-object spectrograph VLT/FLAMES. Here we summarize our most important findings:\\
{\bf -} We found a multi-modal MDF that is nicely reproduced by the combination of four gaussian distributions, in good agreement with \citet{johnson_10}. Our sample is dominated by a MP sub-population, that contributes $44\%$ to the total population. The secondary peaks at higher metallicities contribute to the $35\%$, $14\%$, and $7\%$, respectively.\\
{\bf -} Based on our metallicity distribution, we divided the entire sample into four sub-populations (MP, M-int1, M-int2, MR), which we used to investigate the strength of the (anti-) correlations associated with the MgAl chain. Our analysis revealed a clear Mg-Al anti-correlation, with the shape and extension of the correlation varying significantly with the metallicity of the stars being considered. A clear-cut and discrete Mg-Al anti-correlation is present in all stars with metallicity lower than $\sim -1.3$ dex, while for higher values of [Fe/H], the anti-correlation is less evident or possibly not present at all. \\
{\bf -} We also detected Mg-Si and Al-Si (anti-) correlations, which extensions vary as a function of the metallicity, and as for the Mg-Al anti-correlation, their presence is evident for the stars with [Fe/H] $\lesssim -1.3$ dex. All the observed (anti-) correlations here found confirm the results found in previous works by \citet{NDC95}, \citet{smith_00}, and \citet{meszaros_21}. These results constitute a fingerprint of Mg burning at very high temperatures ($\gtrsim 10^8$ K) through the MgAl chain, at least in the MP and M-int1 sub-populations (\citealt{ventura_16}).  \\
{\bf -} Our Mg-Al anti-correlation presents a discrete shape in the MP and M-int1 sub-populations. In particular, in the MP sub-population, we can observe (1) a Mg-Al anti-correlation analogous to the one observed in genuine single-metallicity GCs (\citealt{meszaros15}; \citealt{pancino17}), with small variations in Mg abundances ($\sim 0.3$ dex) and almost 1 dex of variation in the Al abundances; (2) a distinct component of Mg-poor stars that are all enriched in Al at [Al/Fe] $\sim +1.15$ dex. This kind of sub-population has been observed only among the most metal-poor GCs such as M15, M92, and NGC 5824 (\citealt{masseron_19}; \citealt{mucciarelli_18}), or massive GCs like NGC 2808 (e.g., \citealt{carretta_18}). The Mg-Al anti-correlation in the M-int1 sub-population is dominated by the most Al-rich stars, with a second group of stars at lower Al values and enhanced in Mg. The Al-rich stars in MP and M-int1 sub-populations are clearly separated by $\sim 0.2$ dex, with a gap not detected by \citet{meszaros_21}. \\
{\bf -} In the [Al/Fe] vs [Fe/H] plane we can clearly recognize a trend as a function of the metallicity for the stars with [Al/Fe] $> \sim +0.5$ dex. We can observe that the [Al/Fe] distribution reaches its maximum at [Fe/H] $\sim -1.3$ dex and then there is a decrease of the Al abundances towards the highest metallicities. \\
{\bf -} By comparing our results with those of \citet{meszaros_21} we observe that the fraction of 2P stars decreases from the cluster center towards the outer regions. This finding confirms that the formation of 2P stars is more prevalent in the central regions of the cluster (see \citealt{marino_12} and references therein). The prevalence of 2P in our sample can be the key factor at the origin of all the observed differences between our sample and that of \citet{meszaros_21}.

\subsection{$\omega$ Centauri as a globular cluster}
$\omega$ Centauri exhibits the most extensive chemical anomalies associated with the MgAl chain, making it a unique opportunity to impose additional constraints on the potential nature of the polluters responsible for the multiple populations. 
Indeed, the MgAl chain is far more sensitive to temperature than the CNO and NeNa cycles (\citealt{ventura_16}). Additionally, the presence of stars enhanced in Al (and Si) and depleted in Mg in the MP and M-int1 components of $\omega$ Centauri necessitates the occurrence of proton capture processes at temperatures that cannot be reached in all the polluter stars suggested in the literature.\\
If the polluters responsible for the observed anti-correlations are AGB and super-AGB stars, then the chemical anomalies here observed and the trend of Al abundances with respect to metallicity for stars with [Al/Fe] $> +0.5$ dex can be readily explained. In the metal-poor domain, here represented by the MP and M-int1 populations, the clear Mg-Al trend is due to the strong hot bottom burning (HBB) experienced by low-mass, massive AGB stars, where the ignition of proton-capture nucleosynthesis at temperatures above $10^8$ K favours the depletion of the overall Mg in favour of Al (\citealt{ventura_16}). In this context, the lower peak value of Al exhibited by MP stars with respect to the M-int1 counterparts is due to the activation of the full MgAlSi nucleosynthesis in the most metal-poor AGBs, with the efficient activation of the $^{27}\textrm{Al}$ proton capture reaction, which destroys part of the Al synthesized by Mg burning (\citealt{dellagli_18}). This understanding is confirmed by the Mg-Si and Al-Si trends detected in MP stars. In the M-int2 sub-population the Mg spread is shorter than in the MP and M-int1 sub-populations, since the HBB temperatures experienced by AGB stars of metallicity [Fe/H] $\sim -1$ are not sufficiently hot to favour an extended destruction of the Mg. The lack of a Mg-Si anti-correlation in this sub-population is a signature of inefficiency of the advanced MgAl chain reaction $^{26}\textrm{Al}(p,\gamma)^{27}\textrm{Si}(e^-,\nu)^{27}\textrm{Al}(p,\gamma)^{28}\textrm{Si}$ at high metallicities. Indeed, high metallicities does not allow an efficient Si production. Finally, the short extension of the Mg-Al trend shown up by the most metal-rich stars witnesses the action of proton-capture reactions by the two least abundant $^{25}\textrm{Mg}$ and $^{26}\textrm{Mg}$ isotopes, whereas the HBB temperatures at these metallicities are not sufficiently hot to activate efficiently the proton capture process by the most abundant $^{24}\textrm{Mg}$ isotope: the overall Mg spread is narrow in this case. Therefore, our study shows for the first time the presence in $\omega$ Centauri of the two channels of Al production and destruction (\citealt{ventura_13}; \citealt{dellagli_18}). In conclusion, at [Fe/H]$< \sim -1.3$ dex the Al production channel is always activated, with the destruction channel that becomes significant at the lowest metallicities. On the other hand, at [Fe/H] $> \sim -1.3$ dex, the Al production channel weakens as the metallicity increases, while the destruction channel is not present at all since we do not have any Si production through Al burning. \\
While the AGB model appears to qualitatively account for observed chemical anomalies, it is important to note that different levels of dilution of the AGB ejecta with pristine gas are required to reproduce the observed (anti-) correlations (\citealt{dellagli_18}). Based on their Mg and Al abundance values, 2P stars with less extreme compositions may have formed from AGB ejecta mixed with up to $70\%$ pristine gas, whereas the most extreme populations (characterized by heavy Mg depletion and Al enhancement) may have formed from AGB gas with either very limited or no dilution with pristine material. However, the precise physics and timing of the dilution process during the early evolution of the cluster remain unknown and can possibly require some degree of fine tuning.\\
FRMS (\citealt{krause_13}) or interacting binaries (\citealt{mink_09}) are among the polluter candidates since they are able to activate the CNO cycle and the secondary chains, but they require very high masses (of the order of $\sim 100 M_{\odot}$ or above) and some adjustement of the reaction rates in order to reproduce the observed Mg-Al anti-correlations in GCs (\citealt{prantzos_17}). Supermassive stars ($\sim 1000 M_{\odot};  $\citealt{denissenkov_hart_14}) have central temperatures high enough at the beginning of the main sequence to allow the simultaneous burning of He, Na, and Mg. Moreover, the models show that in these stars Si can be produced, but at temperatures where Mg is heavily destroyed. This is in contrast with what we find in $\omega$ Centauri. Also novae were suggested as polluters (\citealt{maccarone_12}; \citealt{denissenkov_14}). The fundamental issue with stars of this type is that all the light-elements which are enhanced in 2P stars (Na, Al, and Si) are regularly overproduced in quantities that are significantly greater than the reported levels. However, as novae outbursts are multi-parameter phenomena, more research on the parameter space is required to determine the precise amount of light elements ejected by these stars into the intra-cluster medium. \\
Even though many different scenarios have been proposed up to this point, none of them are fully free from serious flaws (e.g., \citealt{renzini_15};~\citealp{renzini_22};~\citealt{bastian_18}, ~\citealt{milone_22}). In particular, all the self-enrichment models cannot explain that 2P stars generally outnumber their 1P counterparts (see \citealt{bastian_15} for a discussion).\\
The available chemical evidence suggest that $\omega$ Centauri chemical enrichment history was very complex  and influenced by a simultaneous contribution of core-collapse supernovae (CC-SNe), as demonstrated by the observed spread in Fe, enhanced [$\alpha$/Fe] ratios, and high [Na/Fe] and [Al/Fe] abundances (see \citealt{johnson_10} and references therein), and likely AGB stars, which are responsible for the observed light-element variations. In contrast to regular (non-nucleated) galaxies or genuine globular clusters, $\omega$ Centauri chemical history has been controlled by its ability to retain both high and low-velocity ejecta. 

\subsection{$\omega$ Centauri as a nuclear remnant}
In the above discussion we have considered our results in the perspective of the origin of the multiple populations in globular clusters \citep{gratton_12,bastian_18,gratton_19}, exploiting the constraints provided by the extreme chemical manifestations of this syndrome that are observed to occur in $\omega$~Centauri. We have also made some attempt to interpret general trends within the entire sample assuming that they are produced by a single chemical evolution path, driven by self-enrichment. However, the latter is just an hypothesis, since, depending on the actual nature of the system, other kind of processes may have been involved in the origin of the present day status of $\omega$~Centauri. In this section want to re-consider the observational scenario from a different perspective.

The idea that $\omega$~Centauri can be the nuclear remnant of a dwarf galaxy whose main body was completely disrupted by the interaction with the Milky Way dates back to decades ago and was the subject of extensive literature \citep[see, e.g.,][and references therein]{gnedin2002,bekki_03,bekki2019}. The strict analogy with the stellar nucleus of the currently disrupting Sgr dSph galaxy was firstly noted and discussed by \citet{bellazz08} and \citet{carretta10}. In recent times the possible association of $\omega$~Centauri with a specific dwarf accretion event, Gaia-Sausage-Enceladus \citep[GSE][]{helmi2018,belokurov2018}, lent further support to this hypothesis \citep{myeong2018,massari2019,linberg2022}. 

There is general consensus that the \citep[widely diffused, see, e.g.][]{boker2004} stellar nuclei are formed by the spiral-in to the center of the host galaxy of massive star clusters, by dynamical friction, and/or by central in situ star formation, with the first channel possibly being the preferred one in $M\la 10^9~M_{\odot}$ galaxies \citep[][and references therein]{nadine20}. Such a multiple-channel formation path can greatly help in accounting for the extremely complex abundance patterns observed in $\omega$~Centauri. 

In this context, we want to highlight two facts that emerges particularly clearly from our analysis and that suggest that indeed the system may be a nuclear remnant that was built up by the merging of globular clusters plus in situ star formation at the center of the (now disrupted) progenitor dwarf galaxy \citep[see also][]{iba2019,iba2021}:

{\bf -} The MDF is clearly multi-modal, with the strongest peak being the most MP one. This is at odd with what observed in local dwarfs \citep{Kirby2011,hassel21}. In these sites, where the build-up of the MDF should be dominated by the chemical evolution of a self-enriching stellar system embedded in a dark matter halo, MDFs have typically a very clean single mode toward the metal-rich side of the distribution plus an extended metal-poor tail. It is interesting to note that this is true also for the Sgr dSph when the MDF is sampled outside the nucleus \citep{muccia2017,minelli_21}, while strong bi-modality emerges in the nuclear region \citep{bellazz08,muccia2017,alfaro-cuello19,alfaro-cuello20}.\\
{\bf -} If we consider the different sub-groups as classified by the GMM described in Sect.~\ref{sec:ferro} we can infer the intrinsic metallicity dispersion with the maximum likelihood analysis described in \citet{muccia2012}, following \citet{pm93} and \citet{walker06}. Doing this, we obtain mean metallicities of the various components in excellent agreement with the results of the GMM, and the intrinsic  metallicity dispersions ($\sigma_{int}$) and the associated uncertainties reported in Table~\ref{tab:sigmet}. It is very interesting to note that both the MP and the M-int1 components are fully consistent with null dispersion, that is the most likely outcome of the analysis, with tiny uncertainty. Null or very small metallicity dispersion is a defining characteristic of globular clusters \citep{ws12,gratton_19}. On the other hand, the M-int2 metallicity distribution is strongly incompatible with zero dispersion and nothing relevant can be said  on the MR population as the uncertainty on $\sigma_{int}$ is huge.\\\\

\begin{table}
      \caption[]{Intrinsic metallicity dispersion of the four sub-populations}
         \label{tab:sigmet}
        \centering          
\begin{tabular}{lc}
\hline\hline \\ [-1.5ex]   
            Group &  $\sigma_{int}$ \\
                        &  (dex)     \\
            \hline
            MP &   $0.00\pm 0.01$   \\
            M-int1 & $0.00\pm 0.02$     \\
            M-int2 & $0.08\pm 0.01$     \\
            MR & $0.00\pm 0.17$     \\
\hline                  
\end{tabular}
\end{table}

Taken at face value these results suggest that $\omega$~Centauri may be indeed the nuclear remnant of a dwarf galaxy that was originally built up by the merging of two massive metal-poor globular clusters (both more massive than $10^6~M_{\odot}$, according to Table \ref{tab:met} and the total mass by \citealt{baumgardt_18}, $M=3.94\pm0.02\times 10^6~M_{\odot}$), each one displaying its own extended light-elements anti-correlations typical of GCs in this mass regime, plus some more metal-rich component, with significant metallicity dispersion, possibly formed in situ, similar to the case of the nucleus of Sgr dSph \citep{carretta2010b,carretta10,alfaro-cuello19,alfaro-cuello20}.

It is important to keep in mind that the reliability of the results reported in Table~\ref{tab:sigmet} depends on the accuracy of the errors on the individual [Fe/H] estimates that is notoriously difficult to assess properly. However, there is little doubt that the intrinsic metallicity dispersion observed in our MP and Mint1 samples, if not null, is very small and compatible with that observed in other massive GCs \citep{carretta2010b,carretta10_1851,lardo_23}, hence the hypothesis that they trace the population of ancient globulars that merged at the center of the progenitor dwarf galaxy 
building the backbone of its stellar nucleus appears sustainable in any case. 

It may be legitimate to ask oneself the reason why the possibility that the M-int1 and, in particular, the MP components can be single-metallicity populations has not emerged so sharply in previous studies. The comparisons between the distribution in various chemical planes suggest that our set of measures has higher precision than comparable samples by other authors, allowing us to get a deeper insight into trends and distributions in these planes (see, e.g., Fig.~\ref{fig:6panel}). As a quantitative test in this sense we considered the sub-samples of MP stars in common with \citet[][170 stars]{johnson_10} and with \citet[][70 stars]{meszaros_21}. Since we are dealing exactly with the same set of stars, the intrinsic metallicity dispersion $\sigma_{int}$ is fixed. When we compute the observed metallicity dispersion $\sigma_{oss}$, in the very reasonable hypothesis of gaussian distribution, $\sigma_{oss}^2 = \sigma_{int}^2 + \sigma_{err}^2$, where $\sigma_{err}$ is the contribution of the measuring error that is different from set of measures to set of measure. The one having the largest $\sigma_{oss}$ will necessarily have also the largest $\sigma_{err}$, hence lower precision. For the MP stars in common with \citet{johnson_10} we obtain $\sigma_{oss}=0.077$~dex from our measures and $\sigma_{oss}=0.137$~dex from their measures.
For the MP stars in common with \citet{meszaros_21} we obtain $\sigma_{oss}=0.085$~dex from our measures and $\sigma_{oss}=0.103$~dex from their measures. Therefore, at least for the considered samples, in both cases our [Fe/H] measures are significantly more precise than theirs. This is clearly a factor enhancing our capability to disentangle the various components of the overall MDF and to properly estimate their metallicity dispersion. The differences in the spatial distribution of the various samples may also play a role, since the different components may have different radial distributions \citep{bellini_09,johnson_10} and our sample is more centrally concentrated than those by \citet{johnson_10} and, especially, by  \citet{meszaros_21}. 

In summary, the scenario outlined above seems to deserve a more thorough and deeper dedicated analysis, taking into account also different aspects of the problem not considered here, like, e.g., the age distribution and the kinematics of the various components. This is clearly beyond the scope of the present paper and we postpone it to a future contribution.

\begin{acknowledgements}
    This research is funded by the project LEGO – Reconstructing the building blocks of the Galaxy by chemical tagging (P.I. A. Mucciarelli). granted by the Italian MUR through contract PRIN 2022LLP8TK\_001. C. Lardo acknowledges funding from Ministero dell'Università e della Ricerca (MIUR) through the Programme {\em Rita Levi Montalcini} (grant PGR18YRML1). We would like to thank Donatella Romano for the useful discussions. 
\end{acknowledgements}

%
   \bibliographystyle{aa} 
   \bibliography{biblio} 
%

\end{document}